\documentclass[12pt]{iopart}

\usepackage{graphicx}

\begin{document}
\begin{flushright}
Preprint {DUKE-TH-02-223}
\end{flushright}

\title{A Relativistic Parton Cascade with Radiation}

\author{Ghi R. Shin\dag\ddag\ and Berndt M\"uller\ddag}
\address{\dag\ Department of Physics, Andong National University,
                    Andong, South Korea}
\address{\ddag\ Department of Physics, Duke University,
                    Durham, NC 27708-0305, USA}

\date{\today}
~ \\

\begin{abstract}
We consider the evolution of a parton system which is formed at
the central rapidity region just after an ultrarelativistic heavy
ion collision. The evolution of the system, which is composed of
gluons, quarks and antiquarks, is described by a relativistic
Boltzmann equations with collision terms including radiation
and retardation effects. The equations are solved by the test
particle method using Monte-Carlo sampling. Our simulations
do not show any evidence of kinetic equilibration, unless the
cross sections are artificially increased to unrealistically large
values.
\end{abstract}

\maketitle

\section{Introduction}
    The possible formation and evolution of a quark-gluon plasma
(QGP) in relativistic heavy-ion collisions has been an important
and very active research subject in recent years. Besides the
theoretical studies of possible signatures of QGP formation, many
investigations have been concerned with space-time dynamics of the
QGP, its formation and final decay into hadrons. Perhaps the most
common description of the evolution of the QGP has been based on
relativistic hydrodynamics \cite{bjo83,baym}. This approach,
however, does not address many important issues, in particular,
the initial formation and local equilibration of the QGP. To
explore this issue theoretically, a microscopic description of the
parton system is required, which allows one to follow the
phase-space evolution in detail. One such approach is the parton
cascade model \cite{gei92,gei95,zpc,mol}.

    An important conceptual problem faced by such microscopic
models is that they require a very detailed specification of the
initial state. Due to our ignorance of nonperturbative processes
in QCD, we do not know how to extract the required information
from the parton wavefunctions of the colliding nuclei. Since
the original formulation of the parton cascade model, important
progress in our understanding of the parton structure of large
nuclei at small Bjorken-$x$ physics has been made \cite{mcl94}.
There is good reason to believe that the parton distribution
at the beginning of a relativistic heavy ion collision can be
described by means of semiclassical methods. This improvement of
our understanding the initial state has motivated a resurgence
of theoretical studies of the microscopic evolution of the
matter formed in the central rapidity region in a nuclear
collisions \cite{mol,bor}.

    To date, those studies have mostly focused on elastic
collisions among partons, although the parton cascade model
was originally formulated to include radiative processes
as time-like and space-like branchings in the leading
logarithmic approximation \cite{gei92}. It has been pointed
recently by Baier et al. \cite{bai00} that the inelastic
$gg \rightarrow ggg$ process is the essential driving
force toward the kinetic equilibration of the parton system.
The importance of this process for the chemical equilibration
of the QGP was already studied and pointed out by Bir\'o et
al. \cite{bir93}.

    There is also a serious defect in many numerical
implementations of the cascade model: the superluminal
propagation of interactions \cite{kor95,zpc,mol}. This becomes 
important whenever the hard-sphere picture of scattering
interactions is applied without proper incorporation of
the retardated nature of interactions in quantum field theory.

    In this article, we develop a cascade code which includes
the process($gg\rightarrow ggg$) as well as a relativistic
picture of scattering that eliminates the superluminal problem.
We assume a boost-invariant initial state and analyze the time
evolution of the parton system. Because we do not include a
model of the hadronization process, our description is useful
only for a few fm/$c$ after the onset of the reaction.

In section 2 we present the equations of motion describing
the evolution of a relativistic system composed of quarks,
antiquarks, and gluons. In section 3 we explain the cascade code, 
especially the choice of the initial state and the parton dynamics. 
Our results and conclusions are discussed in section 4.

\section{Equations of Motion}

    We consider here a system, which consists of gluons, quarks
$(u, d, s)$ and their antiquarks, formed just after the impact of
a heavy ion collision. The typical de Broglie wave length of a
particle in the system, $\lambda = \hbar / pc$, is much shorter
than the average distance between particles even at relatively
high density so that we assume those particles can be considered
as relativistic classical particles. We thus assume that the
semi-classical but relativistic approach is applicable to this
highly energetic and dense parton system. This may not be true for
some particle whose energy is less than 1 GeV, but we limit our
consideration to observables for which the particle picture is
expected to be a good approximation.

    The system which we are studying can be described by
the Lorentz-covariant Boltzmann equation \cite{gro80,cse94},
\begin{eqnarray}
p^\mu \partial_\mu f (x, \bi{p} ) &=& C(\bi{x}, \bi{p}, t),
\label{BTE}
\end{eqnarray}
where $f(x,\bi{p})$ is a phase space distribution function and the
right-hand side denotes the collision terms. We assume that there
is no external force acting on the system and the self-generated
forces of the system are fully described by collisions.

    Gluons and quarks $(u, d, s)$ and their antiparticles $(\bar u,
\bar d, \bar s)$ are independent constituents so that the system
is a mixture of 7 different kinds of partons. These particles can
scatter on each other elastically and they can produce new
particles. We will consider here only the dominant processes among
many possible ones: $g g \leftrightarrow gg$, $g q \leftrightarrow
g q$, $g \bar q \leftrightarrow g \bar q$, $g g \rightarrow g g
g$, $q q \leftrightarrow q q$, $\bar q \bar q \leftrightarrow \bar
q \bar q$, $q \bar q \leftrightarrow q \bar q$, $g g
\leftrightarrow q \bar q$.

Including only those processes among partons, the Boltzmann
equation for gluons takes the form:
\begin{eqnarray}
p^\mu \partial_\mu f_{g} (x, \bi{p} ) &=&  \int_2\int_3\int_4
\frac{1}{2}W_{gg \rightarrow gg} [ f_g(3)f_g(4) - f_g(1)f_g(2) ] 
\nonumber\\
&+& \int_2\int_3\int_4 W_{gq \rightarrow g q} [ f_g(3)f_q(4)
- f_g(1)f_q(2)] \nonumber \\
&+& \int_2\int_3\int_4\int_5 \frac{1}{6}W_{gg \rightarrow ggg} [
f_g(4)f_g(5)- f_g(1)f_g(2) ] \nonumber \\
&+& \int_2\int_3\int_4 W_{g \bar q
\rightarrow g \bar q}[ f_g(3)f_{\bar q}(4) - f_g(1)f_{\bar
q}(2) ] \nonumber\\
&+& \int_2\int_3\int_4 \frac{1}{2}[ W_{q \bar q \rightarrow g g}
f_q(3)f_{\bar q}(4) - W_{g g\rightarrow q \bar q }f_g(1)f_g(2)],
\label{BTE1}
\end{eqnarray}
where we used the abbreviated notations $f_g(i) = f_g(\bi{x}, \bi{p}_i;
t)$, $q =(u, d, s)$, $\bar q = (\bar u, \bar d, \bar s)$, and
$\int_i = \int d \bi{p}_i/E_i$. Note that we assign the
label $1$ to the momentum $\bi{p}$ and set $f_g(1)f_g(2)
\rightarrow f_g(3)f_g(4)f_g(5)$ for loss and $f_g(4)f_g(5)
\rightarrow f_g(1)f_g(2)f_g(3)$ for gain in $gg\rightarrow ggg$
process. We explicitly include the particle symmetry factor in the
classical limit.

We also obtain the evolution equations for quarks and antiquarks,
\begin{eqnarray}
p^\mu \partial_\mu f_q (x, \bi{p} ) &=& \int_2\int_3\int_4 C\,
W_{qq' \rightarrow q q'} [ f_q(3)f_{q'}(4) - f_q(1)f_{q'}(2) ] 
\nonumber \\
&+& \int_2\int_3\int_4 W_{gq \rightarrow g q} [ f_q(3)f_g(4) -
f_q(1)f_g(2) ] \nonumber \\
&+& \int_2\int_3\int_4 \frac{1}{2} [ W_{gg \rightarrow q\bar q}\,
f_g(3)f_g(4) - W_{q\bar q \rightarrow gg}\, f_q(1)f_{\bar q}(2) ].
\label{BTE2}
\end{eqnarray}
\begin{eqnarray}
p^\mu \partial_\mu f_{\bar q}(x, \bi{p} ) &=& \int_2\int_3\int_4
C\, W_{\bar q \bar q' \rightarrow \bar q \bar q'} [ f_{\bar
q}(3)f_{\bar q'}(4) - f_{\bar q}(1)f_{\bar q'}(2) ] \nonumber \\
&+& \int_2\int_3\int_4 W_{g \bar q \rightarrow g \bar q} [ f_{\bar
q}(3)f_g(4) - f_{\bar q}(1)f_g(2) ] \nonumber \\
&+& \int_2\int_3\int_4 \frac{1}{2} [ W_{gg \rightarrow q\bar q}\,
f_g(3)f_g(4) - W_{q\bar q \rightarrow gg}\, f_{\bar q}(1)f_q(2) ].
\label{BTE3}
\end{eqnarray}
where $C = 1/2$ if the final state consist of identical particles
and $C = 1$ otherwise. We neglect in these Boltzmann equations
those quantum mechanical effects coming from Bose enhancement
factors $(1+f_g)$ for a gluon in the final state and Pauli
blocking factors $(1-f_q)$ for final-state quarks or antiquarks.
The quantum transition rates can be expressed in terms of
differential cross sections, e.g.
\begin{eqnarray}
W_{gg_1 \rightarrow g'g_1'} = s \sigma^{gg\rightarrow
gg}(s,\theta) \delta^{(4)}(p+p_1-p'-p_1'),
\end{eqnarray}
where $s$ is the CM energy squared.

These equations are highly non-linear and are impossible to
solve analytically even with simplest initial state. To find
solutions to these equations, we use the test particle method
with Monte-Carlo sampling.

\section{Numerical Simulation}

    The main idea to solve the Boltzmann equations
(\ref{BTE1}, \ref{BTE2}, \ref{BTE3}) is as follows.
We assume that we know the initial state of the system,
the position and momentum of each particle at a given time:
\begin{eqnarray}
f(\bi{x},\bi{p}, t_0) &=& \sum_i^N \delta(\bi{x} - 
\bi{x}_i(t_0))\delta(\bi{p}-\bi{p}_i(t_0)) \label{idis}
\end{eqnarray}
We ignore in our study any collective phenomena which might be
important, and we also ignore the quantum correlations and the
nonperturbative properties of the strong interaction. With these
simplifications, each particle moves freely as time proceeds, but
collides occasionally with another particle if it comes close
enough, i.e., within the distance determined by total cross
section $\sqrt{\sigma_T/\pi}$. In the collision, both particles
abruptly change their momenta according to the differential cross
section and/or produce new particles depending on the branching
ratio of the various final states. The particles emerging from the
collision are again moving freely until they make a collision with
another particle.

    The Monte-Carlo simulation in general cascade code has several
shortcomings:
\begin{enumerate}
\item The deterministic interpretation of the total cross section
as a geometric criterion that decides whether two particles are
colliding with each other or not is not a faithful representation
of the quantum mechanical scattering process. It would be more
appropriate to use a collision probability as a function of the
impact parameter based on a quantum transition amplitude.
\item Another important issue is the assumption on the collision
space-time: The two particles collide each other at their shortest
distance or maximal force point(s). This assumption makes a big
difference between a general (parton) cascade code and an
simplified analytic solution scheme \cite{bor, bai00} or a
relaxation time method \cite{am00, nayak, ser}. The collision in a
cascade code occurs only if there is substantial rapidity
difference and/or transverse momentum difference since the
shortest distance or maximal force point(s) between particles can
be achieved in rather far future if two particles have same
rapidity unless they have relatively high transverse momentum. On
the other hand, most of collisions come from the small angle
scattering between comoving partons in an analytic solution,
namely collisions occurs among those in same rapidity. We believe
this collision condition should be relaxed to allow those
collisions between particles even if they are not in closest
distance.
\end{enumerate}
We hope to address these issues in future publications.

\subsection{Initial Distribution}

It is very difficult to obtain a realistic initial distribution
(\ref{idis}) for the simulation, because of the complexity of nuclear
parton distributions and the dynamics governing the decoherence of
the nuclear wavefunctions. The following choice is mainly governed
by considerations of simplicity and easy implementation in the
cascade code. Note that we are considering only head-on collisions,
although our code is fully three-dimensional.

    As far as the transverse phase space distribution is concerned,
reasonable choices can be derived from the semiclassical picture
\cite{mcl94}. We assume that the number of produced partons is
proportional to the number of primary collisions so that the
transverse spatial distribution of the produced partons has the
imprint of the transverse parton distribution of the colliding
nuclei. This assumption can be expressed as the probability
distribution for the initial transverse position $(x,y)=(r_\perp
\cos\phi, r_\perp \sin\phi)$:
\begin{eqnarray}
P(x,y) = B(1-{\frac{r_\perp^2}{R^2}})
\end{eqnarray}
where $R$ is the radius of the nuclei and $B$ is a normalization
constant. Once the transverse distance from the collision
axis is selected, we can choose the azimuthal angle with equal
probability over $(0, 2\pi)$.

    Krasnitz, Nara and Venugopalan (KNV) \cite{knv00} found that the
transverse momentum distribution from the study of small x-physics
is given by the formula,
\begin{eqnarray}
\frac{1}{\pi R^2}\frac{dN}{dy d^2 p_\perp} &=& \frac{1}{g^2}
f_n(p_\perp/\Lambda_s) \label{KNV1}
\end{eqnarray}
where
\begin{eqnarray}
f_n(p_\perp/\Lambda_s) &= \frac{a_1}
{\exp(\sqrt{p_\perp^2+m^2}/T_{eff}) -1 }, \quad & p_\perp < 3
\Lambda_s , \nonumber\\
&= a_2 \Lambda_s^4 \ln (4\pi p_\perp/\Lambda_s) p_\perp^{-4},
\quad & p_\perp > 3 \Lambda_s
\label{KNV2}
\end{eqnarray}
where $a_1 = 0.0295$, $a_2=0.0343$, $m=0.067\Lambda_s$,
$T_{eff}=0.93\Lambda_s$. $y$ is rapidity and the saturation
momentum $\Lambda_s = 1 GeV$ in RHIC and $2 GeV$ in LHC.

It is known that the rapidity of the produced particles after a
heavy-ion collision is flat in the central rapidity region so
that the rapidity $y$ can be chosen from a constant distribution
within the interval $y_{min} < y < y_{max}$. Once we have chosen
the rapidity, the longitudinal momentum is given by the relation
\begin{eqnarray}
p_z = \sqrt{m^2 + p_T^2} \sinh y.
\end{eqnarray}

We further assume that all of the primary collisions occur in the
reaction plane, $z=0$, at time $t=0$ and the produced partons are
born at the proper time, $\tau_0 = \sqrt{t^2 - z^2}$, where
$\tau_0$ is about 0.3 fm/$c$ at RHIC and 0.13 fm/$c$ at the LHC.
We thus can find the initial time and longitudinal position of a
produced parton once its rapidity is given, by calculating its
velocity $\beta_z = p_z/E$ and setting $\beta_z = z/t$ to obtain
$t$ and $z$ at the formation time $\tau_0$. For simplicity, we
initialize all partons at a common time $t'=\tau_0$ and at
$z'=\beta_z\tau_0$. However, we do not allow the partons interact,
until they reach a proper time $\tau$ larger than the formation
time $\tau_0$.

We assume that the number density per rapidity of produced partons
is in the range $dN/dy \approx 1000 \sim 1500$.

\subsection{Cross Sections}

The dynamics enters into the Boltzmann equation through the total
and differential cross section among partons. Differential cross
sections at leading order $\alpha_s$ for the processes are
stated here explicitly for convenience, although they have been
given elsewhere \cite{izk80},
\begin{eqnarray}
{\frac{d\sigma^{gg\rightarrow gg}}{dt}} &=&
{{9\pi\alpha_s^2}\over{2s^2}}(3-{{tu}\over s^2} - {{su}\over{t^2}}
- {{st}\over{u^2}} ) \label{gg_gg}
\end{eqnarray}
\begin{eqnarray}
{{d\sigma^{gg\rightarrow q_a\bar q_b}}\over{dt}} &=&
{{\pi\alpha_s^2}\over{6s^2}} \delta_{ab} ( {u \over t} + {t \over
u} - {9\over 4}{{t^2+u^2}\over{s^2}} ) \label{gg_qbq}
\end{eqnarray}
\begin{eqnarray}
{{d\sigma^{gq\rightarrow gq}}\over{dt}} &=&
{{4\pi\alpha_s^2}\over{9s^2}} ( - {u\over s} - {s\over u} + {9
\over 4}{{s^2+u^2}\over{t^2}}) \label{gq_gq}
\end{eqnarray}

\begin{eqnarray}
{{d\sigma^{q_aq_b\rightarrow q_aq_b}}\over{dt}} &=&
{{4\pi\alpha_s^2}\over{9s^2}}[{{s^2+u^2}\over{t^2}}
+\delta_{ab}({{t^2+s^2}\over{u^2}}-{2\over 3}{{s^2}\over{ut}})]
\label{qq_qq}
\end{eqnarray}

\begin{eqnarray}
{{d\sigma^{q_a\bar q_b\rightarrow q_c\bar q_d}}\over{dt}} &=&
{{4\pi\alpha_s^2}\over{9s^2}}[ \delta_{ac}\delta_{bd}
{{s^2+u^2}\over{t^2}} + \delta_{ab}\delta_{cd}
{{t^2+u^2}\over{s^2}}-\delta_{abcd}{2\over 3}{u^2\over{st}}]
\label{qbq_qbq}
\end{eqnarray}
\begin{eqnarray}
{{d\sigma^{q_a\bar q_b\rightarrow gg}}\over{dt}} &=&
{{32\pi\alpha_s^2}\over{27s^2}} \delta_{ab} [{u \over t}+{t\over
u}-{9\over 4}{{t^2+u^2}\over{s^2}}]
\label{qbq_gg}
\end{eqnarray}
and
\begin{equation}
{{d\sigma^{g\bar q\rightarrow g\bar
q}}\over{dt}} = {{d\sigma^{gq\rightarrow gq}}\over{dt}},\qquad
{{d\sigma^{\bar q \bar q\rightarrow \bar q \bar q}}\over{dt}} =
{{d\sigma^{qq\rightarrow qq}}\over{dt}}.
\end{equation}

We note that Eq.(\ref{gg_gg}, \ref{gg_qbq}, \ref{qq_qq},
\ref{qbq_gg}) are symmetric under the exchange of $u
\leftrightarrow t$ as dictated by the indistinguishability
of the identical particles. The classical cross sections
must therefore include an overall factor 1/2.

    To get total cross sections, we integrate analytically
the differential cross sections using the conditions, $E_{cm} > 1$
GeV and the minimum momentum transfer $p_\perp > 0.5$ GeV so that
the scattering angle $\theta$ in the CM frame should satisfy the
condition $ |\sin \theta| > (0.5$ GeV$)/E$, where $E$ is the
energy of the particle in CM frame. With this limitation on the
scattering angle, we find $\sigma_{gg \rightarrow gg} \approx 10$/
GeV$^2$ and $\sigma_{qq\rightarrow qq} \approx 0.6$/GeV$^2$ and
$\sigma_{gq\rightarrow gq} \approx 0.7$/GeV$^2$ for the strong
coupling constant $\alpha_s = 0.3$, which we will use throughout
this work. These total cross sections are optimistic values for
those perturbative processes because we choose low values for
minimum CM energy as well as minimum momentum transfer. However,
we note that about similar values for the cross section have been
used by others \cite{nar01, mol}

    We use the following approximation for the $gg \rightarrow ggg$ 
cross section \cite{bir93},
\begin{eqnarray}
{{d\sigma^{gg\rightarrow ggg}}\over{dq_\perp^2 dy dk_\perp^2}} &=&
{{9 C_A\alpha_s^3}\over{2}}
{{q_\perp^2}\over{(q_\perp^2+\mu_D^2)^2}} \nonumber \\
&& \times {{\Theta(k_\perp \lambda_f - \cosh y) \Theta({{\sqrt{s}}}
- k_\perp \cosh y)}\over{k_\perp^2\sqrt{ (k_\perp^2 + q_\perp^2 +
\mu_D^2 )^2 - 4 k_\perp^2 q_\perp^2}}} \nonumber \\
\label{gg_ggg}
\end{eqnarray}
where $q_\perp$ is the momentum transfer of two colliding gluons
and $k_\perp$ the perpendicular momentum of the radiated gluon and
$y$ is its rapidity. $C_A = 3$ for the SU(3) gauge group,
$\mu_D$ is a Debye screening mass, and $\lambda_f$ is the mean
free path. The first step function represents the
Landau-Pomeranchuk-Migdal(LPM) effect while the second one is
the energy constraint of the radiated gluon.

    In order to integrate (\ref{gg_ggg}), we need several
approximations, which we discuss now. For the Debye mass we use
the value that is consistent with the $gg\rightarrow gg$ cross
section:
\begin{eqnarray}
\mu_D^2 = 2\pi C_{gg} \alpha_s^2 / \sigma_T^{gg\rightarrow gg} ,
\end{eqnarray}
where $C_{gg} = 3$. This approximation gives us $\mu_D \approx 0.4$
GeV for $\sigma_{gg\rightarrow gg} = 10$/GeV$^2$. The mean free
path, $\lambda_f$, is defined as $1/(\rho \sigma_T)$ where $\rho$
is the density of particles and $\sigma_T$ is the average total
cross section of the parton. This is a dynamical variable even
if we fix the total cross section, because the system is rapidly
expanding. We can estimate the initial density for heavy ion
collision to be $0.34\sim 0.5$/GeV with $4000 \sim 6000$ gluons
at $\tau_0 = 0.3$ fm/$c$ so that the mean free path is
$0.2 \sim 0.3$/GeV. But this number quickly increases as the system
expands. As a reasonable approximation we choose $\lambda_f = 1.0$/GeV
in our study. With these two parameter choices, we can integrate
the differential cross section numerically. Using the Runge-Kutta
algorithm, we evaluate the total cross section
$\sigma_T^{gg\rightarrow ggg}\approx 3.6$/GeV$^2$

    We can add the total cross section of each process to get the
total cross section, e.g.,
\begin{eqnarray}
\sigma_T^{gg} &=& \sigma_T^{gg\rightarrow gg} +
\sigma_T^{gg\rightarrow ggg} + \sigma_T^{gg\rightarrow u\bar u}
+ \sigma_T^{gg\rightarrow d\bar d}+ \sigma_T^{gg\rightarrow
s\bar s}.
\end{eqnarray}
This total cross section will be used to decide whether two coming
gluons will collide or not, by the criterion whether the impact
parameter between two gluons in the reaction plane is less than
$\sqrt{\sigma_T/\pi}$ or not. Once it is determined that a
collision takes place, we can use the branching ratio to select
which channel will become active. After we choose the channel, we
can use the differential cross section of the channel to select a
scattering angle and momentum transfer.

\subsection{Basic Algorithm}

    We first select the global laboratory frame as the reference
frame in which the simulation is performed. We next consider all
possible pairs of particles and calculate their impact parameter
$b$. The impact parameter is defined as the shortest distance
between two particles in their reaction frame, i.e. their mutual
CM frame with one particle moving in the $+z$ direction and the
other moving in the $-z$ direction. To determine $b$, we apply
an appropriate Lorentz transformation to each pair of particles.
Suppose the particle 1, which has the position $(x_1,y_1,z_1)$,
moves along the +z-axis and particle 2, at $(x_2,y_2,z_2)$, moves
along the $-z$-axis in the reaction frame. The impact parameter is
explicitly given by:
\begin{eqnarray}
b = \sqrt{(x_1-x_2)^2+(y_1-y_2)^2}.
\end{eqnarray}
We check whether a collision is possible by applying the
criterion $b < \sqrt{\sigma_T /\pi}$ as discussed before. We also
check whether the condition $z_1 < z_2$ is met.

    For a given particle, there maybe are many possible collisions
but we keep only the earliest one, as observed in the laboratory
frame. We next calculate where and when two particles will collide.
In order to determine the collision point(s) for two particles, we
make use of the ``maximal force'' principle, which is explained
in detail in Appendix A. In brief, we postulate that the two
particles are interacting through a retarded long-range field.
This assumption is consistent with our use of lowest-order
perturbative cross sections. We determine the point on each
particle's trajectory where the retarded force becomes maximal.
In general, these points correspond to different times in the
laboratory frame, as an expression of the fact that the forces
among particles propagate no faster than the speed of light.

    If the time difference between the two times corresponding
to a single collision is greater than the mean free path, there
is no collision between them, because they will be interrupted by
neighboring particles. This can reduce the number of collisions
substantially as the particle density goes up, or the relative
CM energy of a particle pairs increases. This mechanism is similar
to the procedure A of ref. \cite{kor95} (restriction of the signal 
velocity). Here, however, we do not impose an {\em ad-hoc} cut
on allowable collisions; rather, the finite signal velocity is
embodied in the determination of the interaction points for each
particle (see Appendix A).

    We then put all the collision events in time sequence and
execute the collisions one by one in the global time frame.
After each collision, the collision sequence must be updated,
because the particles emerging from collision have different
phase space coordinates than before.

\subsection{Bulk Properties}

Since we have the full phase space information of all of particles
composing our system, we can, in principle, calculate all the
properties of the system. In particular, we can define the
particle number current and the energy-momentum tensor as follows,
\begin{eqnarray}
N^\mu(x) &=& \int {{d^3p}\over E} p_\mu f(x,\bi{p}),\\
T^{\mu\nu}(x) &=& \int {{d^3p}\over E} p^\mu p^\nu f(x,\bi{p}).
\end{eqnarray}
The classical entropy current is defined as:
\begin{eqnarray}
S^\mu(x) &=& - \int {{d^3p}\over E} p^\mu f(x,\bi{p}) \ln f(x,\bi{p}).
\label{entropy}
\end{eqnarray}
We also find it helpful to calculate the absolute momentum
function weighted by the distribution
\begin{eqnarray}
\tilde{P}_x(\bi{r},t) &=& \int {{d^3p}}\, |p_x| f(x,\bi{p}),\\
\tilde{P}_z(\bi{r},t) &=& \int {{d^3p}}\, |p_z| f(x,\bi{p}) .
\end{eqnarray}
Local kinetic equilibrium requires $\tilde{P}_x = \tilde{P}_z$.

The entropy of the system of classical particles cannot be
calculated from (\ref{entropy}) since the phase space of
the test particles is given by a sum of $\delta$-functions,
\begin{eqnarray}
f(\bi{x},\bi{p}, t) &=& \sum_i \delta(\bi{x}-
\bi{x}_i(t))\delta(\bi{p}- \bi{p}_i(t)),
\end{eqnarray}
where $(\bi{x}_i,\bi{p}_i)$ is the position and momentum of the
$i$-th particle. In order to define a smooth phase space density
we smear this distribution in phase space such that the
uncertainty relation is satisfied. We choose a spatial volume
of extension $\Delta x$ and a momentum space volume of extension
$\Delta p$ such that $\Delta x \Delta p_x > 1$. We use the
representations of the Dirac $\delta$-function as a Gaussian:
\begin{eqnarray}
\delta(x-x_i) &\approx& \sqrt{{a}\over{\pi}} e^{-a(x-x_i)^2}
\label{smear}
\end{eqnarray}
and smooth the probability up to a few nearest momentum bins and
normalize the distribution so that the integrated particle
probability remains unity. This is known as the coarse graining
procedure and the entropy of a system is dependent on the
procedure. In our calculation, we assign the probability,
Eq.(\ref{smear}), up to the third nearest spatial bin from the
position of the particle. Using this method, we calculate the
one-particle entropy to be $3.2$ for an isolated particle.

\section{Results and Discussions}

We are considering a system of which the total number of particles
(gluons) is 6000. The particles have a flat rapidity distribution
between $+2$ and $-2$. The total cross section for the simulation
has been calculated using the coupling constant $\alpha_s = 0.3$
and minimal momentum transfer 0.5 GeV as described before. We
also include the effect of higher order diagrams in the form
of a K-Factor $K=2$. We use the KNV transverse momentum distribution
({\ref{KNV1},\ref{KNV2}) for the $p_T$ distribution.

    Figure 1 shows total number of particles as function of time.
The primary partons are initialized at the proper time $\tau_0$
($=1.5$/GeV at RHIC energy), and the secondary partons are
produced by collisions between primary partons of the system.
In the following, when we use the term ``produced partons'',
we refer to those partons that are created in collisions within
our transport code, but not to the initialized partons that are
produced before the start of our transport calculation.
With our parameters, the secondary partons, mostly gluons,
are not produced as abundantly as predicted in the analytical
estimate of Baier, Mueller, Schiff and Son \cite{bai00}.
One reason for this difference is that the cross section
(\ref{gg_ggg}) is small at early times due to the influence
of the LPM effect, which provides for a strong suppression
of radiation at high density.
\begin{figure}
\includegraphics{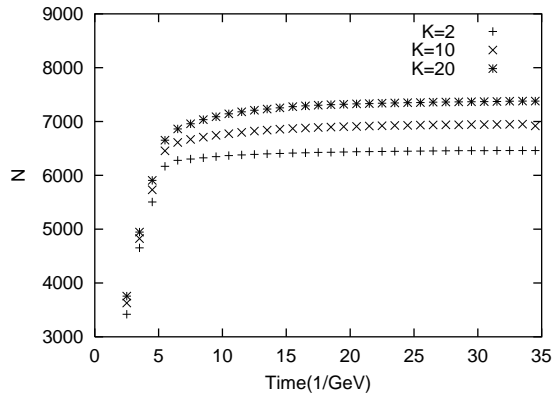}
\caption{Total number of partons as a function of time in 1/GeV
unit: 6000 initial gluons and $K = 2, 10, 20$. Data are
average over 20 runs which use different random number sets.}
\label{fig1}
\end{figure}
\begin{figure}
\includegraphics{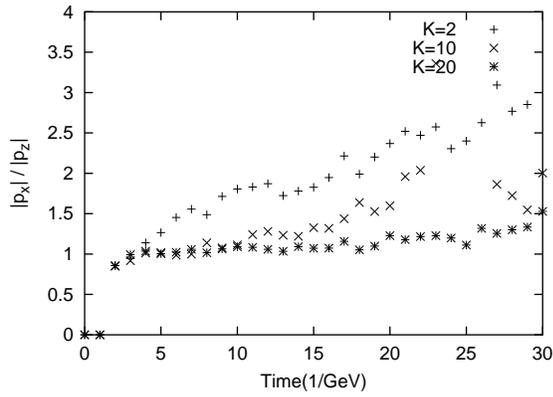}
\caption{$\sum{|p_x|}/\sum{|p_z|}$ as a function of time in 1/GeV
unit: 6000 initial gluons and K-Factor = 2(+), 10(x) and 20(*).
Data are the average over 20 runs in which each run uses different
set of random numbers. A small box ($2\times 2\times 1 fm^3$) at
center is used for measurement.}\label{fig2}
\end{figure}

    Figure 2 shows the ratio between transverse momentum and
longitudinal momentum of the particles in a small box, which
has the size of $2\times 2\times 1$ fm$^3$ and is located at
the coordinate origin, as a function of time. This ratio should
be close to unity when the kinetic equilibration is reached.
The figure shows that the ratio is increasing beyond unity,
implying that the system is thinning more rapidly along the
collision axis than in the transverse direction for $K=2$ and
even for $K=10$. Kinetic equilibration can be achieved for
the unrealistically large cross section with $K=20$.

    Figure 3 shows the energy density of the box as a function
of time. The high energy particles escape from the box very
quickly and a relatively slow expansion follows.
\begin{figure}
\includegraphics{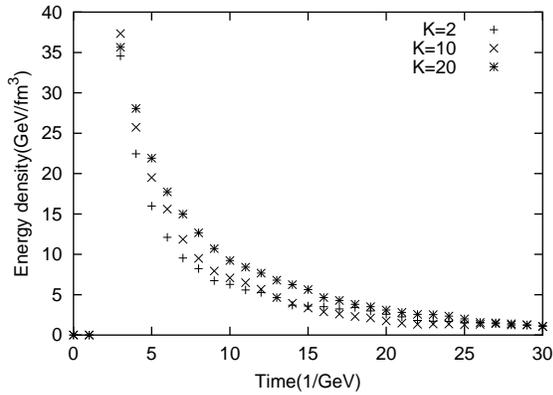}
\caption{The energy density per fm$^3$ in the box at central
region as a function of time .} \label{fig3}
\end{figure}
\begin{figure}
\includegraphics{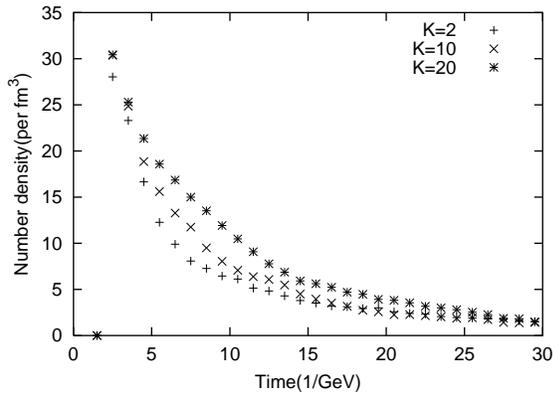}
\caption{The number density per fm$^3$ in the box at central
region as a function of time .} \label{fig4}
\end{figure}

    Figure 4 shows the particle number density of the box which
we are considering as a function of time. We can see that more and
more particles are staying in the box as the cross section gets
larger, but the densities are almost the same after those high
momentum particles escape. Scattering occurs only occasionally
as time goes on.

In conclusion, we do not observe kinetic equilibration for a
realistic cross section in our simulations. However, when we
increase the cross section substantially ($10-20$ times!)
approximate isotropy of the momentum distribution can be 
achieved. In our investigation here, we kept the low-momentum 
fixed. On the other hand, dynamical screening in an expanding
medium provides an effetive low-momentum cutoff that varies with
time. This raises the question whether a selfconsistent treatment,
in which the time-dependent density is used to determine the
time variation of the total cross section, can exhibit an approach
to equilibrium without unrealistic assumptions. We hope to return
to this question, as well as to the issue of including $3 \to 2$ 
processes to ensure detailed balance, in a future publication.

\ack
G.~R.~Shin thanks the members of the Department of
Physics at Duke University for their warm hospitality during his
sabbatical visit. We both thank S.~A.~Bass, D.~K.~Srivastava, and
S.~Mrowczynski for helpful discussions and comments. This work has
been supported in part by DOE grant DE-FG02-96ER40945 and by Andong
National University.

\appendix

\section{Collision points}

Quantum mechanics does not allow us to determine the precise
moment of when a collision between two particles occurs due
to the uncertainty relation. Here, we discuss this question
in the context of classical mechanics, in particular, in
classical electrodynamics. Consider two particles with charges
$q_1, q_2$, and positions and velocities, $(x^\mu,u^\mu)$ and
$(r^\nu, v^\nu)$, respectively. The electromagnetic field
induced by particle 2 at the position of particle 1 is given
by \cite{jac74}
\begin{eqnarray}
F^{\mu\nu}(x) &=& {e\over{[v\cdot(x-r(\tau_x'))]^3}} [
(x-r(\tau_x'))^\mu v^\nu \\ \nonumber & &- (x-r(\tau_x'))^\nu
v^\mu ]
\end{eqnarray}
where $r(\tau_x')$ is the source point associated with the field
at point $x$. Each particle moves along its trajectory at constant
velocity:
\begin{eqnarray}
x^\mu &=& a^\mu + u^\mu \tau ,\\
r^\mu &=& b^\mu + v^\mu \tau'.
\end{eqnarray}
Using the retardation condition, we can obtain the proper time of the
source point for the field point $x$:
\begin{eqnarray}
\tau_x' &=& (x-b)\cdot v - \sqrt{ [(x-b)\cdot v]^2 -(x-b)^2}.
\end{eqnarray}
The force on particle 1 due to the field of particle 2 is given by
\begin{eqnarray}
F^{\mu\nu}u_\nu = {{e[(x-r')^\mu v\cdot u - (x-r')\cdot u v^\mu]}
\over {[[(a-b+u\tau)\cdot v]^2 - (a-b+u\tau)^2]^{3/2}}}
\end{eqnarray}
where $r' = r(\tau_x')$. This force will be maximal when the
denominator is minimal. This condition will be met at
\begin{eqnarray}
\tau_0 &=& {{(a-b)\cdot v (u\cdot v) -(a-b)\cdot u}\over{u^2
-(u\cdot v)^2}}.
\end{eqnarray}
Even though the classical collision occurs gradually, we associate
this space-time point with the collision moment of the particle 1:
\begin{eqnarray}
x^\mu &=& a^\mu + u^\mu \tau_0.
\end{eqnarray}
Using the same procedure, we obtain the maximal force on particle
2 by the particle 1 at the proper time
\begin{eqnarray}
\tau_0' &=& {{(b-a)\cdot u (v\cdot u) -(b-a)\cdot v}\over{v^2
-(u\cdot v)^2}}
\end{eqnarray}
and the collision space-time point
\begin{eqnarray}
r^\mu &=& b^\mu + v^\mu \tau_0'.
\end{eqnarray}
Since the definitions of $\tau_0$ and $\tau_0'$ are manifestly
Lorentz invariant, the interaction points are independent of the
reference frame in which the two-body collision is calculated.

The force between the two particle is more complex due to the
presence of color degrees of freedom. However, in lowest-order
perturbation theory, the space-time behavior of the QCD force
is identical to that in electrodynamics. We can thus apply the
equations derived above.

\section{Monte Carlo Sampling}

    We explain here the Monte-Carlo sampling for the process $gg
\rightarrow gg$ only. The other processes are the same except the
process $gg \rightarrow ggg$ which we discuss briefly at the end.
We change the variable from the Mandelstam $t$ to $\eta=\cos
\theta$. The differential cross section becomes
\begin{eqnarray}
{{d\sigma^{gg\rightarrow gg}}\over{d\eta}} &=&
{{9\pi\alpha_s^2}\over{4E_{cm}^2}} [ 3-{{(1-\eta)(1+\eta)}\over 4}
\nonumber\\ && +2{{1+\eta}\over{(1-\eta)^2}} +2
{{1-\eta}\over{(1+\eta)^2}} ]
\end{eqnarray}
where $\eta$ is confined by the minimal momentum transfer $Q_0$
which we will choose either $0.5$ or $1 GeV$. The total cross
section can be obtained analytically by integrating over the
allowed range. The natural way to choose $\eta$ by monte Carlo
method is that we calculate
\begin{eqnarray}
r &=& \int_{\eta_{min}}^\eta {{d\sigma}\over{d\eta}}
\end{eqnarray}
and invert this equation to solve for $\eta$. Since this is not
easy to do, we divide the $(\eta_{min},\eta_{max})$ into
several intervals. Each interval can be integrated over
analytically to give an area. We generate a random number $r_1$
and select the interval depending on the area, i.e., if $r_1 <
s_1/\sigma_T$, the interval 1 is chosen, and so forth. Once we
have selected the interval $(\eta_{start},\eta_{end})$, we can find
the straight line connecting the points at $\eta_{start}$ and
$\eta_{end}$ and choose $\eta$ between $(\eta_{start},\eta_{end})$
by linear random sampling and then use the accept-reject method.
This triple Monte Carlo sampling gives a satisfactory result.

    The sampling of the $gg \rightarrow ggg$ process is slightly
more complicated. Using the step function in (\ref{gg_ggg}), we find
the rapidity of the bremsstrahlung gluon, $(y_{min}, y_{max})$,
where for a given CM energy $\sqrt{s}$,
\begin{eqnarray}
y_{max} &=& \ln \left ( \sqrt{{{\sqrt{s}\lambda_f}}} +
\sqrt{{{\sqrt{s}\lambda_f - 1}}}\right ).
\end{eqnarray}
Once we have chosen the rapidity, we have
\begin{eqnarray}
\left ({{\cosh y}\over{\lambda_f}} \right )^2 < k_\perp^2 < \left(
{{\sqrt{s}}\over{\cosh y}}\right )^2
\end{eqnarray}
and
\begin{eqnarray}
0 < q_\perp^2 < s.
\end{eqnarray}
Again, we can use the accept-reject method to sample $(k_\perp^2,
q_\perp^2)$. This algorithm generates an acceptable distribution
within a reasonable CPU time.

\end{document}